# Temporal Coupled Mode Theory for a Single Floquet-Sheet Resonator

Yao-Ting Wang and Hsu-Hui Chou

*Department of Photonics, National Sun Yat-sen University, Kaohsiung, 80424, Taiwan*

**Abstract**

We develop a rigorous Temporal Coupled-Mode Theory (TCMT) specifically tailored for a single Floquet-sheet resonator governed by time-modulated conductivities. By invoking photon-number conservation during frequency conversion, we derive characteristic radiative decay rates and coupling coefficients that account for the frequency ratio between channels. We establish a systematic bridge to the Floquet Transfer Matrix Method (TMM), providing closed-form analytical expressions that map scattering parameters to the Drude-type physics of the sheet. Our model explicitly captures the resonant coupling between the $0^{th}$-order propagating channel and the $-1^{st}$-order surface-mode channel. Validated by COMSOL numerical simulations, the theory remains robust even when intrinsic material loss is incorporated. This framework offers an intuitive pole-expansion representation for designing time-varying photonic interfaces.

## I. Introduction

In the latest decade, the field of photonics has witnessed a major transition from spatially inhomogeneous structures to temporally modulated systems. Time-varying materials, often referred to as temporal metamaterials [1,2] or four-dimensional optics, [3] have opened up unprecedented avenues for controlling light-matter interactions, such as temporal aiming [4], multiple light conversion [5], and more [6-12]. By modulating the constitutive parameters of a medium, on a timescale comparable to the optical cycle, researchers have demonstrated exotic phenomena including adiabatic frequency conversion [13], topological effects in photonic time crystals [14], and the electronic circuit for time-reversal wavefronts [15]. While conventional metamaterial/photonic crystals rely on spatial periodicity to control the wavevector, time-varying media utilize temporal modulations to redistribute energy across the frequency spectrum. This dynamic control provides a fertile ground for developing next-generation optical components, such as non-reciprocal isolators [16] and efficient periodic synthetic space in frequency domain [17-20], which are essential for both optical communications and light signal processing.

In the recent years, the exploration of time-varying effects has naturally narrowed down from three-dimensional bulk media to two-dimensional interfaces, where the extreme sub-wavelength confinement of light leads to enhanced light-matter interaction. Recent works have established the basic principle for time-varying metasurfaces or Floquet sheets. For instance, studies by Galiffi et. al. have elucidated the scattering physics at time-modulated boundaries [22], highlighting how temporal variations in surface impedance can trigger surface plasmon polariton excitations and anomalous frequency shifts. Building upon these concepts, researchers investigated symmetry-selective surface-mode excitation in double Floquet sheets [23], where the relative phase difference between temporal modulations serves as a critical degree of freedom to efficiently switch between even and odd plasmonic modes. Similarly, research by Wang et. al [24] demonstrated how space-time metasurfaces can break reciprocity and realize extreme non-reciprocity without the need for bulky magnetic components. These developments suggest that time-varying interfaces are not merely simplified models of bulk media but are unique platforms that facilitate strong coupling between propagating modes and localized surface states via the coupling between Floquet harmonics.

To provide a unified and analytically tractable description of these resonant systems, TCMT has emerged as a cornerstone in the photonic community. TCMT was firstly proposed by Fan and colleagues [25-27] to describe systems with energy conservation and time-reversal symmetry. In their a series of seminal articles, a resonator is characterized by its eigenfrequency, radiative decay rates, and coupling coefficients to external ports. For time-modulated systems, TCMT can be used as a powerful tool to separate different modal coupling. It allows for the description of frequency conversion as a resonant process, where energy is exchanged between different Floquet channels through a discrete set of coupling equations, ensuring that the complex physics of time-varying systems can be understood through a compact, intuitive pole-expansion representation.

Despite the success of TCMT in describing various time-varying photonic resonators such as Ref. [28], its application to Floquet-sheet resonators, specifically those governed by time-modulated conductivities/Drude weight, remains unexplored. While numerical approach like transfer matrix method (TMM) can simulate these systems, a rigorous analytical bridge that connects the macroscopic scattering parameters to the microscopic physics of the sheet is still missing. Particularly, the relationship between the surface modal excitation in the -$1^{st}$ Floquet channel and the resulting scattering in the $0^{th}$ channel has not been fully formulated within a TCMT framework. This paper seeks to fill this gap by developing a TCMT model specifically tailored for a single Floquet sheet and establishing a systematic connection to the TMM. Furthermore, the underlying concepts and the mapping procedure established herein are generalizable to a wide range of other time-varying resonant systems.

In this paper, we develop a TCMT description for a single Floquet-sheet resonator and establish a systematic connection to a Floquet transfer-matrix formulation. The content of this article is organized as follows: In Sec. II, we present a TCMT model for the resonator and derive the general scattering response in terms of a background contribution and a resonant term, while postponing explicit closed-form expressions for key parameters, such as radiative decay rates, to later sections. Sec. III provides a brief review of the conducting thin layer, discussing the fundamental properties and conventional theoretical treatments relevant to this study. In Sec. IV, we formulate the corresponding TMM for the Floquet sheet and derive analytical expressions that map the TMM quantities to all TCMT parameters, thereby providing a closed-form expression retrieval procedure. In Sec. V, we incorporate intrinsic loss and use the resulting theory to produce absorption spectrum, yielding explicit design conditions and validating them within the same formalism. Finally, Sec. VI summarizes our main results and discusses possible extensions, including multi-sheet configurations and more general Floquet-channel coupling scenarios.

## II.   TCMT for a Single Floquet Sheet

Figure 1 illustrates a Floquet sheet whose surface current density obeys a lossless Drude-type dynamic equation $dJ/dt = \eta_0^{-1} W_p(t) E_x$, where $\eta_0$ is vacuum impedance and the Drude weight $W_p(t)$ is temporally modulated as

$$W_p(t) = W_{p0}(1 + 2\alpha \cos \Omega t). \tag{1}$$

If the surface modal excitation is described by a complex amplitude $a$, then the excitation dynamics induced by the Floquet sheet with modulation frequency W can be written in the coupled-mode form [29]

$$\frac{da}{dt} = (-i\omega_s + \gamma_s)a + e^{i\Omega t}\boldsymbol{\kappa}^T \mathbf{s}_+, \tag{2}$$

where $\omega_s$ is the eigenfrequency of the surface mode, $\gamma_s$ accounts for radiative leakage of the surface mode into the external scattering channels, and $\boldsymbol{\kappa}$ denotes the coupling coefficient between the incident waves and the surface excitation. The amplitude is normalized so that $|a|^2$ equals the energy stored in the surface mode. In the ideal lossless limit considered here, the intrinsic dissipation rate $\gamma_i$ is set to be zero. The presence of material losses will be discussed in the subsequent sections. In addition, the radiation from the Floquet sheet back into the propagating channels is described by

$$\mathbf{s}_- = C\mathbf{s}_+ + e^{-i\Omega t}\mathbf{d}a \qquad (3)$$

where $C$ is the direct $S$-matrix pertaining to the incoming and outgoing waves of the sheet, and $\mathbf{d}$ characterizes how the resonant surface excitation radiates into the external channels. The explicit time factors in Eq. (2) and (3) are introduced to satisfy the frequency matching condition to connect the input channel at $\omega_0$ and the surface mode channel at $\omega_{-1} = \omega_0 - \Omega$.

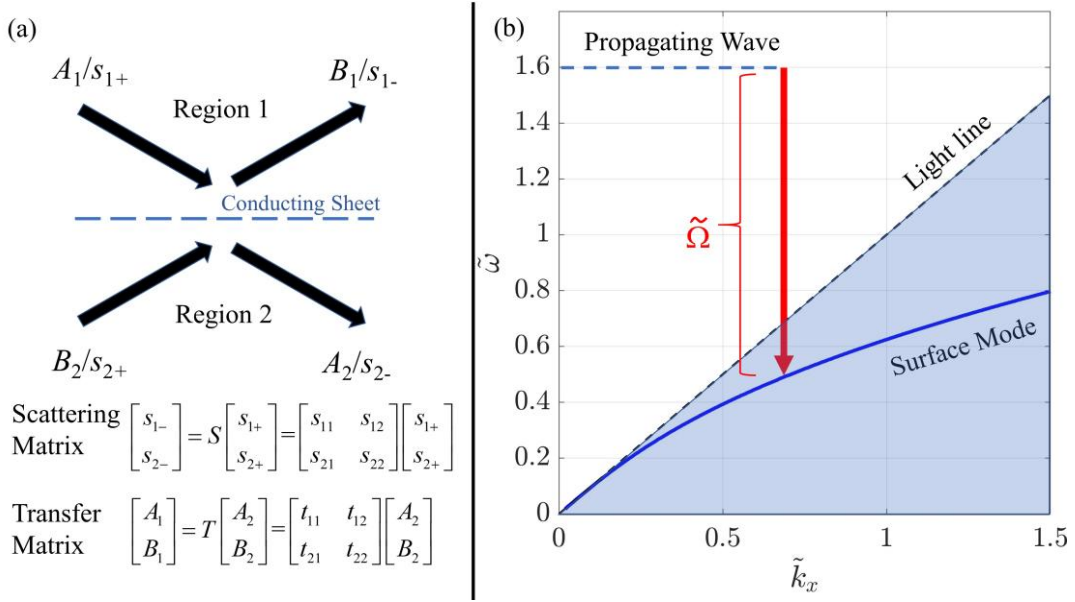

Fig. 1: (a) Schematic of the incident and outgoing wave components across the conducting sheet in region 1 and region 2. The corresponding scattering matrix $S$ and transfer matrix $T$ definitions are provided below. (b) Normalized dispersion diagram ($\tilde{\omega} = \omega/W_{p0}$ and $\tilde{k}_x = ck_x/W_{p0}$) illustrating the light line, the supported surface mode, and the frequency modulation process $\tilde{\Omega}$ coupling a propagating wave to the surface state.

When the incident wave $\mathbf{s}_+$ impinges on the Floquet sheet at a frequency $\omega_0$, the time modulation enables frequency conversion between the propagating channels at $\omega_0$ and the surface-mode channel centered around $\omega_{-1} = \omega_0 - \Omega$ so that the detuning entering the resonance denominator is $\omega_0 - \Omega - \omega_s$. Solving Eq. (2) and (3) in the frequency domain then yields the scattering relation

$$\mathbf{s}_- = S\mathbf{s}_+ = \left[ C + \frac{\mathbf{d}\boldsymbol{\kappa}^T}{\gamma_s - i(\omega_0 - \Omega - \omega_s)} \right]\mathbf{s}_+. \qquad (4)$$

To determine the radiative decay rate $\gamma_s$ and the coupling vector $\mathbf{d}$ under our chosen normalization of the surface-mode amplitude, we invoke photon-number conservation for the frequency-conversion process. Although the sheet is temporally modulated and electromagnetic energy is not required to be conserved, a monochromatic conversion event still allows one to enforce conservation of the number of photons exchanged between the surface excitation at $\omega_s$ and the radiated waves at $\omega_0$. With the normalization $|a|^2$ representing the energy stored in the surface mode, we define the surface-mode photon number as $N_s = |a|^2/\omega_s$. For the propagating channels, the photon flux is proportional to $N_0 = |\mathbf{s}|^2/\omega_0$. Hence, photon-number conservation requires

$$\frac{d}{dt}\left(\frac{|a|^2}{\omega_s}\right) = \frac{|\mathbf{s}_-|^2 - |\mathbf{s}_+|^2}{\omega_0}. \tag{5}$$

The left-hand side represents the change of surface modal photon numbers in time, whereas the right-hand side is the photon numbers after the interaction with single Floquet-sheet resonator. Substituting Eq. (2) and its time-conjugate into the left-hand side of Eq. (5), it becomes

$$\frac{d}{dt}\left(\frac{|a|^2}{\omega_s}\right) = \frac{2\gamma_s |a|^2 + e^{i\Omega t} a^* \boldsymbol{\kappa}^T \mathbf{s}_+ + e^{-i\Omega t} a \boldsymbol{\kappa}^\dagger \mathbf{s}_+^*}{\omega_s}. \tag{6}$$

Next, from Eq. (3), the intensity of the outgoing wave can be written as

$$|\mathbf{s}_-|^2 = |C|^2 |\mathbf{s}_+|^2 + |\mathbf{d}|^2 |a|^2 + e^{i\Omega t} a^* \mathbf{d}^\dagger C \mathbf{s}_+ + e^{-i\Omega t} a \mathbf{s}_+^\dagger C^\dagger \mathbf{d}. \tag{7}$$

In the lossless background S-matrix, $|C| = 1$. By plugging Eq. (3) into the right-hand side of Eq. (5) yields

$$\frac{|\mathbf{s}_-|^2 - |\mathbf{s}_+|^2}{\omega_0} = \frac{|\mathbf{d}|^2 |a|^2 + e^{i\Omega t} a^* \mathbf{d}^\dagger C \mathbf{s}_+ + e^{-i\Omega t} a \mathbf{s}_+^\dagger C^\dagger \mathbf{d}}{\omega_0}. \tag{8}$$

Equating Eq. (6) and (8) for arbitrary $(a, \mathbf{s}_+)$ then fix the coefficients uniquely

$$2\gamma_s = (\omega_s / \omega_0) |\mathbf{d}|^2, \tag{9a}$$

$$\boldsymbol{\kappa}^T / \omega_s = \mathbf{d}^\dagger C / \omega_0. \tag{9b}$$

The factor $\omega_s / \omega_0$ is the characteristic correction arising from photon-number conservation across frequency conversion: photon flux scales as $|\mathbf{s}|^2 / \omega_0$, so matching the decay of $N_s$ to the net radiated photon flux necessarily introduces the ratio of the surface-mode and radiated-wave frequencies.

Building on the above conservation of photon numbers, we next derive additional constraints on the coupling structure imposed by time-reversal symmetry and reciprocity, in direct analogy with the relations obtained in static two-port resonators. To this end, we consider the time-reversed counterpart of the radiative-decay scenario. Starting from the leaky surface-mode solution with no incident wave ($\mathbf{s}_+ = 0$) and an outgoing wave $\mathbf{s}_-(t) = e^{-i\Omega t}\mathbf{d}a(t)$, time reversal maps $(a, \mathbf{s}_+, \mathbf{s}_-)$ to $(a^*, \mathbf{s}_-^*, \mathbf{s}_+^*)$. Therefore, the time-reversed solution corresponds to a coherent incidence $\hat{T}_R \mathbf{s}_+(t) = \mathbf{s}_-^*(-t)$ that produces no outgoing wave, i.e., $\hat{T}_R \mathbf{s}_- = \mathbf{s}_+^*(-t) = 0$. Substituting this condition into time-conjugate of Eq. (3) yields the constraint $C\mathbf{s}_-^*(-t) + e^{-i\Omega t}\mathbf{d}a^*(-t) = 0$, which, together with $\mathbf{s}_-^*(-t) = e^{-i\Omega t}\mathbf{d}^* a^*(-t)$, leads to the general time-reversal relation

$$C\mathbf{d}^* + \mathbf{d} = 0, \tag{10}$$

up to an overall phase convention of the modal amplitude.

Combining Eqs. (4), (9), and (10), the total scattering matrix can be written in the following form:

$$S = \left[ C - \frac{(\omega_s/\omega_0)\mathbf{d}\mathbf{d}^T}{\gamma_s - i(\omega_0 - \Omega - \omega_s)} \right]. \tag{11}$$

In conventional TCMT, the coupling coefficient $\mathbf{d}$ and the radiative decay rate $\gamma_s$ are often treated as phenomenological parameters and are obtained by fitting measurable spectra, such as reflection and transmission. Determining $\mathbf{d}$ and $\gamma_s$ analytically, however, requires an additional full-wave description of the structure. In the next section, we start from the TMM to establish the physical equivalence between TCMT and TMM, and to derive closed-form expressions for $\mathbf{d}$ and $\gamma_s$.

### III. Brief Review for Unmodulated Conducting Thin Layer

To establish the TMM framework for a Floquet sheet, we first review the standard electromagnetic results for an unmodulated conducting thin layer, which is subsequently treated as an infinitesimal sheet in the limit where the layer thickness $L$ vanishes. We consider a planar conducting layer characterized by a Drude-type dielectric form:

$$\varepsilon_D(\omega) = \varepsilon_\infty - \frac{\omega_p^2}{\omega(\omega + i\gamma_i)}, \tag{12}$$

where $\gamma_i$ and $\omega_p$ denote the intrinsic loss rate and the plasma frequency, respectively. For a TM-polarized electromagnetic field ($E_x, E_z, H_y$) propagating inside the conducting layer, the transverse component of **E** and **H** field are given by

$$E_{t,D}(z) = A_D e^{ik_{z,D}z} + B_D e^{-ik_{z,D}z}, \tag{13a}$$

$$H_{t,D}(z) = \left(A_D e^{ik_{z,D}z} - B_D e^{-ik_{z,D}z}\right)/\eta_{TM}, \tag{13b}$$

where $k_{z,D} = \sqrt{\varepsilon_D \mu (\omega/c)^2 - k_x^2}$ and $\eta_{TM} = k_{z,D}/\omega \varepsilon_0 \varepsilon_D$. By defining the fields at the boundaries as $E_{t1} = E_{t,D}(0)$, $H_{t1} = H_{t,D}(0)$, $E_{t2} = E_{t,D}(L)$, $H_{t2} = H_{t,D}(L)$, and invoking the relationship $J_{sj} = H_{tj}$, we obtain the surface current densities on both interfaces:

$$J_{s1} = \frac{\eta_S E_{t1} - \eta_T E_{t2}}{\eta_S^2 - \eta_T^2}, \tag{14a}$$

$$J_{s2} = \frac{\eta_S E_{t2} - \eta_T E_{t1}}{\eta_S^2 - \eta_T^2}. \tag{14b}$$

The surface impedances $\eta_S$ and $\eta_T$ are defined as follows:

$$\eta_S = i\eta_{TM} \cot(k_{z,D}L), \tag{15a}$$

$$\eta_T = i\eta_{TM} \csc(k_{z,D}L). \tag{15b}$$

In the limit where the thickness of the layer is much smaller than the wavelength $k_{z,D}L \ll 1$, we employ the approximations $\cot(x) \approx 1/x$ and $\csc(x) \approx 1/x$. Consequently, the total surface current density can be expressed as

$$J_{s1} + J_{s2} \approx -i\omega L \varepsilon_0 \left( \varepsilon_\infty - \frac{\omega_p^2}{\omega(\omega + i\gamma_i)} \right) \left( \frac{E_{t1} + E_{t2}}{2} \right). \tag{16}$$

As $L \to 0$, the sum of surface current densities on both interfaces converges to the net surface current density on a single conducting sheet, while $(E_{t1} + E_{t2})/2$ represents the local average field $E_{t,avg}$. By defining $J_s = \sigma_g E_{t,avg}$, Eq. (16) can be reformulated as

$$\varepsilon_{\text{thin layer}} = \varepsilon_\infty + i\frac{\sigma_g}{\omega \varepsilon_0 L}. \tag{17}$$

This expression is in rigorous agreement with previous results reported in ref. [30]. In the regime where the background permittivity $\varepsilon_\infty$ is dominated by the plasma term, the Drude-type surface conductivity simplifies to

$$\sigma(\omega_0) = \frac{\eta_0^{-1} W_{p0}}{\gamma_i - i\omega_0}. \tag{18}$$

Here, $W_{p0}$ denotes the unmodulated Drude weight as defined in Eq. (1). Again, we emphasize that $\omega_0$ refers the frequency on the zeroth-order Floquet channel rather than resonance frequency.

In the following analysis, we consider the sheet to be situated at the interface between two homogeneous media with permittivity $\varepsilon_j$ and permeability $\mu_j$ ($j = 1, 2$). For an obliquely incident plane wave with a fixed in-plane wavevector $k_x$, the transfer matrix relating the field amplitudes across the sheet is given by

$$T_{0,0} = \frac{1}{2} \begin{bmatrix} 1 + \dfrac{K_0^{(2)} + \Xi_{0,0}}{K_0^{(1)}} & 1 - \dfrac{K_0^{(2)} - \Xi_{0,0}}{K_0^{(1)}} \\ 1 - \dfrac{K_0^{(2)} + \Xi_{0,0}}{K_0^{(1)}} & 1 + \dfrac{K_0^{(2)} - \Xi_{0,0}}{K_0^{(1)}} \end{bmatrix}. \tag{19}$$

The explicit form of $K_0^{(2)}$ and $\Xi_{0,0}^{(2)}$ are given below

$$K_0^{(j)} = \frac{\omega_0 \varepsilon_0 \varepsilon_j}{k_{z,0}^{(j)}}. \tag{20a}$$

$$\Xi_{0,0} = \frac{\eta_0^{-1} W_{p0}}{\gamma_i - i\omega_0}. \tag{20b}$$

From Eq. (19) and (20), the reflection and transmission coefficients for the $0^{\text{th}}$-order channel are

$$r_0 = \frac{K_0^{(1)} - K_0^{(2)} - \Xi_{0,0}}{K_0^{(1)} + K_0^{(2)} + \Xi_{0,0}}, \tag{21}$$

and

$$t_0 = \frac{2K_0^{(1)}}{K_0^{(1)} + K_0^{(2)} + \Xi_{0,0}}. \tag{22}$$

Equation (19) can also be recast into the corresponding scattering-matrix form. Using the standard transfer-to-scattering conversion for a 2-by-2 transfer matrix $T$, i.e., $c_{11} = t_{21}/t_{11}$, $c_{21} = 1/t_{11}$, $c_{12} = \det(T)/t_{11}$, and $c_{22} = -t_{12}/t_{11}$, one obtains

$$C = \begin{bmatrix} r_0 & t_0' \\ t_0 & r_0' \end{bmatrix} = \frac{1}{\Delta_0} \begin{bmatrix} K_0^{(1)} - K_0^{(2)} - \Xi_{0,0} & 2K_0^{(2)} \\ 2K_0^{(1)} & -K_0^{(1)} + K_0^{(2)} - \Xi_{0,0} \end{bmatrix}, \tag{23}$$

where the common denominator is $\Delta_0 = K_0^{(1)} + K_0^{(2)} + \Xi_{0,0}$. In addition to the background scattering described above, the sheet can also support a surface mode. At the surface-mode frequency $\omega_s$, the in-plane wavevector $k_x$ lies outside the light cone, so the normal wavevector becomes purely imaginary such that $k_z^{(j)} = i\kappa_z^{(j)} = i\sqrt{k_x^2 - \omega_s^2 \varepsilon_j \mu_j / c^2}$, leading to an evanescent field away from the interface. For identical media on both sides, the surface-mode dispersion can be obtained from the pole condition of the scattering response, which reduces to $\Delta_0 = 0$. Thus, we have

$$\omega_s = \frac{\eta W_{p0}}{\sqrt{8}} \sqrt{\sqrt{1 + \left(\frac{4k_x c_0}{\mu W_{p0}}\right)^2} - 1}. \tag{24}$$

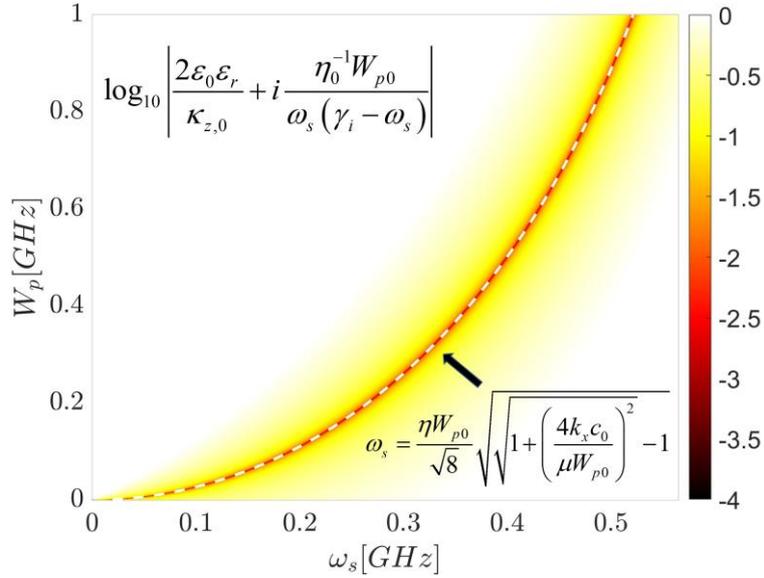

Fig. 2: The colormap illustrates the numerical evaluation of the surface mode derived from the absolute value shown at the top. To enhance the clarity of the mode trajectory, the intrinsic loss term is set to a small value. The dashed white line, obtained from the analytical expression for $\omega_s$ (bottom right), demonstrates excellent agreement with the numerical results, effectively validating the simplified model in the low-frequency regime.

As shown in Fig. 2, the dispersion of the surface mode follows a nonlinear trajectory as the Drude weight $W_p$ increases. The close matching between the colormap and the analytical curve confirms that the simplified pole condition accurately captures the essential physics of the surface-plasmon resonance in this system.

## IV. Determine $d$ and $\gamma_s$ via TMM in Floquet Sheet

With the results of an unmodulated sheet in hand, we now turn to the temporally modulated case. On each Floquet-channel index $n$, we consider a TM-polarized electromagnetic field ($E_x, E_z, H_y$) in a two-region geometry and expand the $x$ component of **E** field in each region as

$$E_{x,n}^{(j)}(z) = A_{j,n} e^{ik_{z,n}^{(j)} z} + B_{j,n} e^{-ik_{z,n}^{(j)} z}. \tag{25}$$

Here $j$ labels the two homogeneous regions on two sides of the conducting sheet. The coefficients $A_{j,n}$ ($B_{j,n}$) corresponds to the forward- and backward-propagating waves, respectively. The frequency in the n-th Floquet channel is $\omega_n = \omega + n\Omega$ and the corresponding normal wavevector component is wavevector $k_{z,n}^{(j)} = \sqrt{\varepsilon_j \mu_j \omega_n^2 / c^2 - k_x^2}$. From the Ampère-Maxwell's law, the associated magnetic field component is

$$H_{y,n}^{(j)}(z) = \frac{\omega_n \varepsilon_0 \varepsilon_j}{k_{z,n}^{(j)}} \left( A_{j,n} e^{ik_{z,n}^{(j)} z} - B_{j,n} e^{-ik_{z,n}^{(j)} z} \right). \tag{26}$$

Based on Eq. (21), the time-modulated conducting sheet is described by a lossy Drude-type dynamic equation for the surface current density,

$$dJ_{x,n}/dt + \gamma_i J_{x,n} = \eta_0^{-1} W_p(t) E_{x,n}^{(2)}, \tag{27}$$

where $W_p(t)$ is the Drude weight given by Eq. (1). Placing the sheet at $z = 0$, and applying the boundary conditions for the transverse fields, one gets a set of coupled equations between Floquet channels. In the following, we focus on the reduced model constraining only $0^{th}$ and -$1^{st}$ channel.

$$\underbrace{\begin{bmatrix} D_0^{(1)} & 0_{2\times 2} \\ 0_{2\times 2} & D_{-1}^{(1)} \end{bmatrix}}_{D_1} \begin{bmatrix} A_{1,0} \\ B_{1,0} \\ A_{1,-1} \\ B_{1,-1} \end{bmatrix} = \underbrace{\begin{bmatrix} D_0^{(2)} + X_{0,0} & X_{0,-1} \\ X_{-1,0} & D_{-1}^{(2)} + X_{-1,-1} \end{bmatrix}}_{D_2} \begin{bmatrix} A_{2,0} \\ B_{2,0} \\ A_{2,-1} \\ B_{2,-1} \end{bmatrix}. \quad (28)$$

The block-matrix elements are defined as

$$D_{0(-1)}^{(j)} = \begin{bmatrix} 1 & 1 \\ K_{0(-1)}^{(j)} & -K_{0(-1)}^{(j)} \end{bmatrix}, \quad K_{0(-1)}^{(j)} = \frac{\omega_{0(-1)} \varepsilon_0 \varepsilon_j}{k_{z,0(-1)}^{(j)}}, \quad (29)$$

and

$$X_{0,0(-1,-1)} = \begin{bmatrix} 0 & 0 \\ \Xi_{0,0(-1,-1)} & \Xi_{0,0(-1,-1)} \end{bmatrix}, \quad \Xi_{0,0(-1,-1)} = \frac{\eta_0^{-1} W_{p0}}{\left(\gamma_i - i\omega_{0(-1)}\right)}, \quad (30a)$$

$$X_{0,-1} = \begin{bmatrix} 0 & 0 \\ \Xi_{0,-1} & \Xi_{0,-1} \end{bmatrix}, \quad \Xi_{0,-1} = \frac{\alpha \eta_0^{-1} W_{p0}}{\left(\gamma_i - i\omega_0\right)}, \quad (30b)$$

$$X_{-1,0} = \begin{bmatrix} 0 & 0 \\ \Xi_{-1,0} & \Xi_{-1,0} \end{bmatrix}, \quad \Xi_{-1,0} = \frac{\alpha \eta_0^{-1} W_{p0}}{\left(\gamma_i - i\omega_{-1}\right)}. \quad (30c)$$

Across the Floquet sheet, the transfer matrix is $T = D_1^{-1} D_2$, and the corresponding block matrices are

$$T_{-1,-1} = \begin{bmatrix} \dfrac{K_{-1}^{(1)} + K_{-1}^{(2)} + \Xi_{-1,-1}}{2K_{-1}^{(1)}} & \dfrac{K_{-1}^{(1)} - K_{-1}^{(2)} + \Xi_{-1,-1}}{2K_{-1}^{(1)}} \\ \dfrac{K_{-1}^{(1)} - K_{-1}^{(2)} - \Xi_{-1,-1}}{2K_{-1}^{(1)}} & \dfrac{K_{-1}^{(1)} + K_{-1}^{(2)} - \Xi_{-1,-1}}{2K_{-1}^{(1)}} \end{bmatrix}, \quad (31a)$$

$$T_{0,-1} = \frac{\Xi_{0,-1}}{2K_0^{(1)}} \begin{bmatrix} 1 & 1 \\ -1 & -1 \end{bmatrix}, \quad (31b)$$

$$T_{-1,0} = \frac{\Xi_{-1,0}}{2K_{-1}^{(1)}} \begin{bmatrix} 1 & 1 \\ -1 & -1 \end{bmatrix}, \quad (31c)$$

and $T_{0,0}$ has been given in Eq. (19).

Moreover, we introduce the state vectors $\mathbf{u}_{j,n} = [A_{j,n}, B_{j,n}]^T$. For region I and II, we further define the extended 4-by-1 state vectors by stacking the Floquet channel

$n = 0$ and $n = -1$ as $\mathbf{v}_j = [\mathbf{u}_{j,0}^T, \mathbf{u}_{j,-1}^T]^T$. The transfer matrix connecting $\mathbf{v}_1$ and $\mathbf{v}_2$ can be written as

$$\begin{bmatrix} \mathbf{u}_{1,0} \\ \mathbf{u}_{1,-1} \end{bmatrix} = \begin{bmatrix} T_{0,0} & T_{0,-1} \\ T_{-1,0} & T_{-1,-1} \end{bmatrix} \begin{bmatrix} \mathbf{u}_{2,0} \\ \mathbf{u}_{2,-1} \end{bmatrix}, \tag{32}$$

Next, because the -1 Floquet channel gives rise to a surface mode, its field must be evanescent away from the sheet, Writing the -1 channel surface field in each region as $E_{x,surf}^{(j,-1)}(z) = A_{j,-1}e^{-k_z z} + B_{j,-1}e^{+k_z z}$, the decaying solutions require the exponentially growing component to vanish in each half-space: in Region I ($z < 0$), naturally we set $A_{1,-1} = 0$, whereas in Region II ($z > 0$), we set $B_{2,-1} = 0$ on the $z > 0$ region. These conditions can be compactly enforced by introducing two row vectors $\mathbf{L}_1^T \mathbf{u}_{1,-1} = 0$ and $\mathbf{L}_2^T \mathbf{u}_{2,-1} = 0$ such that $\mathbf{L}_1^T = [1,0]$ and $\mathbf{L}_2^T = [0,1]$. Applying these constraints to Eq. (32) yields the matrix equation

$$\underbrace{\begin{bmatrix} \mathbf{L}_1^T T_{-1,-1} \\ \mathbf{L}_2^T \end{bmatrix}}_{M, 2\times 2} \mathbf{u}_{2,-1} = -\underbrace{\begin{bmatrix} \mathbf{L}_1^T T_{-1,0} \\ \mathbf{0} \end{bmatrix}}_{N, 2\times 2} \mathbf{u}_{2,0}. \tag{33}$$

Solving Eq. (33) leads to

$$\mathbf{u}_{2,-1} = -M^{-1} N \mathbf{u}_{2,0}. \tag{34}$$

Substituting Eq. (34) into $\mathbf{u}_{1,0} = T_{0,0} \mathbf{u}_{2,0} + T_{0,-1} \mathbf{u}_{2,-1}$, we have

$$\mathbf{u}_{1,0} = \left( T_{0,0} - T_{0,-1} M^{-1} N \right) \mathbf{u}_{2,0} \to T_{\text{eff}} \mathbf{u}_{2,0}. \tag{35}$$

Equation (35) defines an effective (reduced) transfer-matrix-description where the contribution of the -1 Floquet channel is eliminated and incorporated into the $0^{\text{th}}$ channel. Substituting Eq. (31b-c) into Eq. (35), we obtain the explicit expression of $T_{0,-1} M^{-1} N = \delta [1,1;-1,-1]$ where $\delta$ is

$$\delta = \frac{\left( \alpha \eta_0^{-1} W_{p0} \right)^2}{2 K_0^{(1)} \Delta_{-1} \left( \gamma_i - i\omega_0 \right) \left( \gamma_i - i\omega_{-1} \right)} \tag{36}$$

and $\Delta_{-1} = K_{-1}^{(1)} + K_{-1}^{(2)} + \Xi_{-1,-1}$. To excite surface mode, the frequency conversion must reach the dispersion region below light cone. Thus, one can make the propagating wavevector purely imaginary, i.e., $k_{z,-1}^{(j)} \to i\kappa_{z,-1}^{(j)}$. Note that we choose the $+i$ branch to ensure that the surface wave occurs in positive frequency regime. Defining a factor $\beta_{-1}(k_x;\varepsilon_1,\varepsilon_2) = \varepsilon_2 \kappa_{z,-1}^{(1)}/\varepsilon_1 \kappa_{z,-1}^{(2)}$, Eq. (36) becomes a more compact form

$$\delta = \frac{\left[\alpha^2 \eta_0^{-1} W_{p0}/2K_0^{(1)}(\gamma_i - i\omega_0)\right]i\omega_s}{\gamma_i - i(\omega_{-1} - \omega_s)}. \tag{37}$$

In Eq. (37), the surface-mode frequency $\omega_s$ is defined as

$$\omega_s(\omega_{-1},k_x;\varepsilon_1,\varepsilon_2) = \frac{\eta_1 W_{p0}}{1+\beta_{-1}}\sqrt{\left(c^2 k_x^2/\omega_{-1}^2 \varepsilon_1 \mu_1\right)-1}. \tag{38}$$

It is worth note that, when the materials on both sides on are identical, the $\beta$ factor becomes unity and the pole condition $\omega_{-1} = \omega_s$ will yield the functional form of surface-mode frequency shown in Eq. (24).

The resulting scattering coefficients, including the contribution from -1 channel, are

$$s_{11} = r = \frac{K_0^{(1)} - K_0^{(2)} - \Xi_{0,0} + 2K_0^{(1)}\delta}{K_0^{(1)} + K_0^{(2)} + \Xi_{0,0} - 2K_0^{(1)}\delta}, \tag{39a}$$

$$s_{22} = r' = -\frac{K_0^{(1)} - K_0^{(2)} + \Xi_{0,0} - 2K_0^{(1)}\delta}{K_0^{(1)} + K_0^{(2)} + \Xi_{0,0} - 2K_0^{(1)}\delta}, \tag{39b}$$

$$s_{12} = t' = \frac{2K_0^{(2)}}{K_0^{(1)} + K_0^{(2)} + \Xi_{0,0} - 2K_0^{(1)}\delta}, \tag{39c}$$

$$s_{21} = t = \frac{2K_0^{(1)}}{K_0^{(1)} + K_0^{(2)} + \Xi_{0,0} - 2K_0^{(1)}\delta}. \tag{39d}$$

Rewriting Eq. (39) then gives (See Appendix B for the derivation)

$$r = r_0 + \left[\frac{(r_0 + t_0)2K_0^{(1)}}{3K_0^{(1)} - K_0^{(2)} - \Xi_{0,0}}\right]\frac{2K_0^{(1)}\delta}{\Delta_0 - 2K_0^{(1)}\delta} \tag{40a}$$

$$r' = r'_0 + \left[\frac{(r'_0 + t'_0)2K_0^{(1)}}{-K_0^{(1)} + 3K_0^{(2)} - \Xi_{0,0}}\right]\frac{2K_0^{(1)}\delta}{\Delta_0 - 2K_0^{(1)}\delta} \tag{40b}$$

$$t' = t'_0 + \left[\frac{(r'_0 + t'_0)2K_0^{(1)}}{-K_0^{(1)} + 3K_0^{(2)} - \Xi_{0,0}}\right]\frac{2K_0^{(1)}\delta}{\Delta_0 - 2K_0^{(1)}\delta} \tag{40c}$$

$$t = t_0 + \left[ \frac{(r_0 + t_0) 2K_0^{(1)}}{3K_0^{(1)} - K_0^{(2)} - \Xi_{0,0}} \right] \frac{2K_0^{(1)}\delta}{\Delta_0 - 2K_0^{(1)}\delta} \tag{40d}$$

Importantly, so far there are no symmetry constraints required. Despite the complexity of $\delta$ and $\Delta_0$, the entire effect of the excited surface mode is solely governed by the pole structure of $\Delta_0 - 2K_0^{(1)}\delta$. However, Eqs. (40) can be greatly simplified by setting the intrinsic loss term $\gamma_i = 0$, $\varepsilon_1 = \varepsilon_2$, and $\mu_1 = \mu_2$. With $K_0^{(j)} = 2\eta_0^{-1}\eta_j^{-1}/\cos\theta_i$, where $\theta_i$ is the incident angle, we have (See Appendix C for the derivation)

$$S = \begin{bmatrix} r_0 & t_0 \\ t_0 & r_0 \end{bmatrix} - \frac{\gamma_s (r_0 + t_0)}{\gamma_s - i(\omega_{-1} - \omega'_s)} \begin{bmatrix} 1 & 1 \\ 1 & 1 \end{bmatrix}, \tag{41}$$

where

$$\gamma_s = \left( \frac{\omega_s}{\omega_0} \right) \frac{2\alpha^2 \eta_1^{-1} W_{p0} / \cos\theta_i}{\left( 2\eta_1^{-1}/\cos\theta_i \right)^2 + \left( W_{p0}/\omega_0 \right)^2}, \tag{42a}$$

$$\omega'_s = \omega_s \left[ 1 - \frac{\alpha^2 (W_{p0}/\omega_0)^2}{\left( 2\eta_1^{-1}/\cos\theta_i \right)^2 + \left( W_{p0}/\omega_0 \right)^2} \right]. \tag{42b}$$

Comparing Eq. (41) with (11) leads to the results of $\mathbf{d}$ and $\boldsymbol{\kappa}$:

$$\mathbf{d} = \sqrt{\frac{\omega_0}{\omega_s}} \sqrt{(r_0 + t_0)\gamma_s} \begin{bmatrix} 1 \\ 1 \end{bmatrix}, \tag{43a}$$

$$\boldsymbol{\kappa} = \sqrt{\frac{\omega_s}{\omega_0}} \sqrt{(r_0 + t_0)\gamma_s} \begin{bmatrix} 1 \\ 1 \end{bmatrix}. \tag{43b}$$

It is noteworthy that Eq. (41) recovers the functional form of the seminal temporal coupled-mode theory proposed by Fan et. al. [25], This correspondence is underpinned by the $2\pi$ phase difference between the reflection coefficient $r_0$ and transmission coefficient $t_0$, i.e., $r_0 + t_0 = e^{i\phi}(|r_0| + i|t_0|)$, under the condition of identical surrounding materials. This structural identity demonstrates how the surface modal resonance within the -1$^{\text{st}}$ Floquet channel facilitates resonant scattering into the 0$^{\text{th}}$ channel. Furthermore, the radiative rate $\gamma_s$ shown in Eq. (42a) explicitly includes the frequency conversion factor $\omega_s/\omega_0$, providing a macroscopic manifestation of the photon number conversion enforced in the previous section. Eq. (42a) also quantitively links the square of the modulation strength $\alpha^2$, Drude weight $W_{p0}$, and the incident angle $\theta_i$ with the radiative rate, thereby elucidating the physical origin of the resonance in temporally modulated conducting sheet. Additionally, Eq. (42b) reveals that the surface-mode frequency $\omega'_s$ undergoes a subtle shift due to the temporal modulation. To ensure a well-defined resonance with minimal frequency deviation, maintaining a small $W_{p0}/\omega_0$ ratio is essential, which necessitates that the modulation frequency $\Omega$ remains sufficiently higher than the excitation. Finally, the symmetry of the coupling vector **d** and **k** in Eq. (43) confirms that the surface excitation maintains an even distribution when the system possesses mirror symmetry, which is a direct consequence of the mirror symmetry inherent in a sub-wavelength Floquet-sheet resonator. Also, the factors of **d** and **k** evidence the frequency conversion when the mode couples across $0^{th}$ (propagating mode) and $-1^{st}$ (surface mode) channel, which aligns the photon-number conversation in Sec. I.

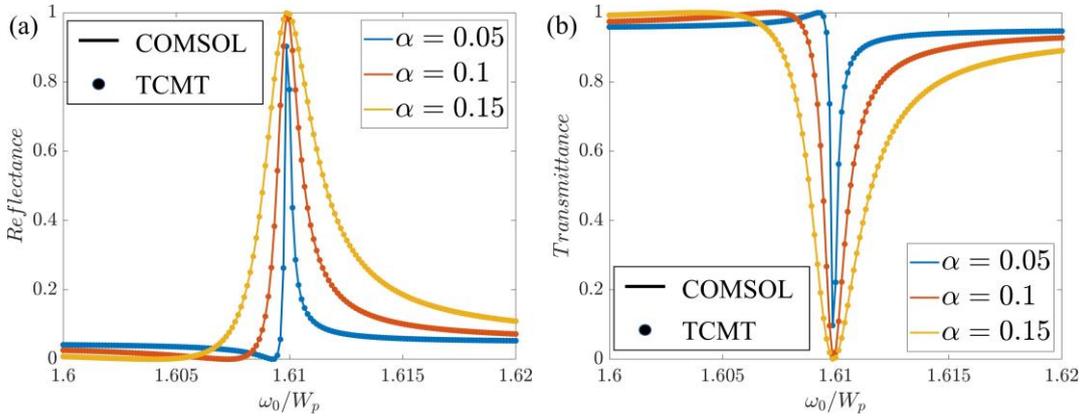

Fig. 3: (a) Reflectance and (b) transmittance as a function of the normalized frequency $\omega_0/W_p$ for various modulation strengths $\alpha = 0.05$, 0.1, and 0.15. The solid lines represent the numerical results obtained from COMSOL Multiphysics, while the dot markers denote the analytical results derived from the TCMT.

In Fig. 3, it is evident that the numerical results from COMSOL and the analytical predictions from the TCMT model exhibit an excellent fit across the entire frequency range, which validates the TCMT model. As the modulation strength $\alpha$ increases, the resonance peak in reflectance and the corresponding dip in transmittance broaden, indicating that stronger temporal modulation enhances the radiation loss and strengthens the coupling between the surface mode and the incident free-space wave.

## V. Absorption for a Single Floquet-Sheet with Intrinsic Loss

In the previous section, we established the TCMT framework for a single Floquet-sheet resonator and determined its key parameters using the TMM. Here, we extend the model to include the intrinsic loss term $\gamma_i$. Based on Eq. (40), the scattering coefficients are modified as follows:

$$r = r_0 - t_0 \frac{\Gamma(\gamma_i, \omega_0)}{\gamma_i - i(\omega_{-1} - \omega_s) + \Gamma(\gamma_i, \omega_0)}, \tag{44a}$$

$$r' = r'_0 - t'_0 \frac{\Gamma(\gamma_i, \omega_0)}{\gamma_i - i(\omega_{-1} - \omega_s) + \Gamma(\gamma_i, \omega_0)}, \tag{44b}$$

$$t' = t'_0 - t'_0 \frac{\Gamma(\gamma_i, \omega_0)}{\gamma_i - i(\omega_{-1} - \omega_s) + \Gamma(\gamma_i, \omega_0)}, \tag{44c}$$

$$t = t_0 - t_0 \frac{\Gamma(\gamma_i, \omega_0)}{\gamma_i - i(\omega_{-1} - \omega_s) + \Gamma(\gamma_i, \omega_0)}, \tag{44d}$$

where $\Gamma(\gamma_i, \omega_0)$ reads

$$\Gamma(\gamma_i, \omega_0) = \left(\frac{\omega_s}{\omega_0}\right) \frac{\alpha^2 \eta_0^{-1} \omega_{p0}}{(1 + i\gamma_i/\omega_0)\Delta_0}. \tag{45}$$

While Eqs. (44) share a similar functional form with Eqs. (40), they exhibit two major discrepancies: the complex nature of both $\Gamma(\gamma_i, \omega_0)$ and $\Delta_0$, and the structural difference between the two rank-one coupling matrices (specifically, the coefficients $r_0 + t_0$ versus $t_0$). The complex nature arises from the violation of photon-number conservation owing to the material absorption. In principle, the functional form of Eqs. (44) cannot be strictly simplified into the resonance term $\gamma_s/[\gamma_i + \gamma_s - i(\omega_{-1} - \omega_s)]$. However, given that the intrinsic loss $\gamma_i$ is negligible compared to the operating frequency $\omega_0$, this approximation remains viable within a wide range of $\gamma_i$. To illustrate the regime in which this approximation remains valid, Fig. 4 shows the reflectance as a function of $\gamma_i$ and $\omega_0$ in vicinity of the resonance. From $\gamma_i/W_p = 10^{-4}$ to $10^{-2}$, both results are in excellent agreement, thereby indicating the use of simple resonance form

$$S \approx \begin{bmatrix} r_0 & t_0 \\ t_0 & r_0 \end{bmatrix} - \frac{\gamma_s(r_0+t_0)}{\gamma_i+\gamma_s-i(\omega_{-1}-\omega'_s)}\begin{bmatrix} 1 & 1 \\ 1 & 1 \end{bmatrix} \qquad (46)$$

is in general valid when $\gamma_i/W_p$ is smaller than 0.01.

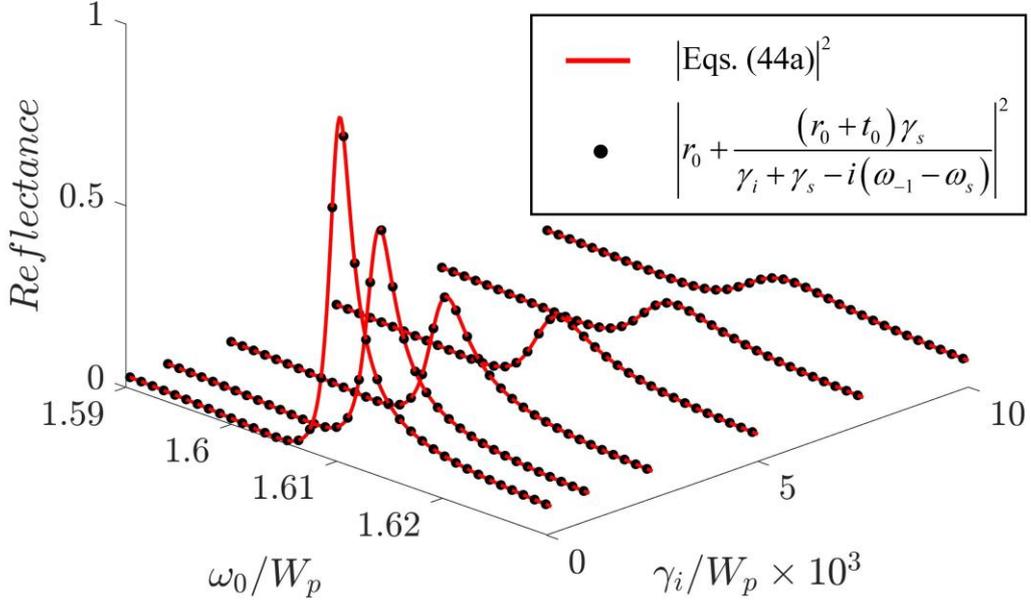

Fig. 4: The reflectance as a function of the normalized frequency $\omega_0/W_p$ and normalized intrinsic loss $\gamma_i/W_p$. The solid red lines represent the numerical results obtained from the full matrix formalism (Eqs. 44a), while the black dots denote the simplified analytical expression shown in the legend. The excellent agreement across all slices validates the robustness of the analytical model in predicting the resonance behavior and linewidth broadening as the material loss increases.

## VI.   Conclusion

In conclusion, we have developed a comprehensive TCMT description for a single Floquet-sheet resonator and established a systematic connection to a Floquet transfer-matrix formulation. To determine the key parameters of the resonator, we introduce photon-number conservation for the frequency-conversion process, which necessitates a characteristic frequency ratio correction in the coupling coefficients to ensure physical consistency. This structural identity demonstrates how the surface modal resonance within the -1$^{st}$ Floquet channel facilitates resonant scattering into the 0$^{th}$ channel, effectively recovering the functional form of seminal TCMT while successfully incorporating the complexities of temporal modulation.

The analytical expressions derived via TMM provide a quantitative link between the modulation strength, Drude weight, and incident angle with the radiative decay rate, thereby elucidating the microscopic physical origin of the resonance in time-modulated-sheet systems. Our investigation into the effects of intrinsic material loss confirms that the simplified resonance form remains robust, with analytical predictions matching full-wave numerical results across a wide range of loss parameters. Furthermore, the symmetry of the coupling vectors confirms that the surface excitation maintains an even distribution due to the inherent mirror symmetry of the sub-wavelength sheet. The mapping procedure and underlying concepts established in this work are generalizable to a broad spectrum of time-varying resonant systems. These results provide a vital theoretical foundation for future explorations into multi-sheet configurations, non-reciprocal topological interfaces, and advanced periodic synthetic spaces in the frequency domain.

**Appendix A: Time-Reversal Invariance of a Single Floquet Sheet**

In this Appendix we justify that a single, lossless Floquet sheet with a purely co-sinusoidal modulation defined as Eq. (1) is invariant under time reversal operation. The sheet enters Maxwell's equations only through a transition boundary condition that relates the jump of the tangential magnetic field to the surface current density, $\hat{\mathbf{n}} \times [\mathbf{H}_1(\mathbf{r},t) - \mathbf{H}_2(\mathbf{r},t)] = \mathbf{J}_s(\mathbf{r},t)$, together with a local surface current relation between $\mathbf{J}_s(\mathbf{r},t)$ and the tangential electric field $\mathbf{E}_\parallel(\mathbf{r},t)$. Under the time-reversal operation $\hat{T}_R$, the electromagnetic fields transform as $\hat{T}_R \mathbf{E}(r,t) \to \mathbf{E}(r,-t)$ and $\hat{T}_R \mathbf{H}(r,t) \to -\mathbf{H}(r,-t)$, while the surface current transforms as $\hat{T}_R \mathbf{J}(r,t) \to -\mathbf{J}(r,-t)$. Applying time-reversal operator to the boundary condition thus yields

$$\hat{\mathbf{n}} \times \left[\mathbf{H}_1(\mathbf{r},-t) - \mathbf{H}_2(\mathbf{r},-t)\right] = \mathbf{J}_s(\mathbf{r},-t), \tag{A1}$$

i.e., the boundary condition retains the same form after time reversal operation. Therefore, the only potential source of time-reversal non-invariance would be the explicit time dependence of the sheet parameter. However, because the modulation is purely co-sinusoidal, it is an even function of time:

$$W_p(-t) = W_{p0}\left[1 + 2\alpha \cos(-\Omega t)\right] = W_p(t). \tag{A2}$$

Hence the sheet response parameter that is controlled by $W_p(t)$ is also invariant under time-reversal operation. Note that, although the modulation may include an arbitrary initial phase, one can always choose the time origin such that this phase is set to zero, without loss of generality. Combining the evenness of the modulation with the standard field transformations under $\hat{T}_R$, we conclude that the time-reversed fields satisfy the same Maxwell equations in the two half-spaces and the same transition boundary condition at the sheet as the original fields. In other words, the time-reversal conjugate system coincides with the original system for a lossless Floquet sheet driven by Eq. (1), establishing time-reversal invariance of a single Floquet-sheet resonator.

**Appendix B: The Derivation of Eq. (40)**

In Appendix B, we derive Eq. (40a) from Eq. (39a). The same procedure is applicable to other equations in Eqs. (39). Based on the identity

$$\frac{Q_1 + Q_2 \delta}{P_1 + P_2 \delta} = \frac{Q_1}{P_1} + \frac{(P_1 Q_2 - Q_1 P_2)\delta}{P_1(P_1 + P_2 \delta)}, \tag{B1}$$

Eq. (39a) can be written as

$$s_{11} = r_0 + t_0 \frac{2 K_0^{(1)} \delta}{\left(\Delta_0 - 2 K_0^{(1)} \delta\right)} \tag{B2}$$

with $\Delta_0 = K_0^{(1)} + K_0^{(2)} + \Xi_{0,0}$. Using some straightforward algebra, Eqs. (21) and (22), one can then obtains Eq. (40a).

**Appendix C: The Derivation of Eq. (41)**

In Appendix C, we derive Eq. (41) from Eq. (40a) as an exemplar, as other elements can be obtained in the same manner. We start from the intermediate step in Eq. (40a)

$$\left(\frac{2K_0^{(1)}}{3K_0^{(1)}-K_0^{(2)}-\Xi_{0,0}}\right)\frac{2K_0^{(1)}\delta}{\Delta_0-2K_0^{(1)}\delta}$$

$$=\frac{\left(\dfrac{2K_0^{(1)}}{3K_0^{(1)}-K_0^{(2)}-\Xi_{0,0}}\right)\dfrac{\alpha^2\eta_0^{-1}W_{p0}i\omega_s}{(\gamma_i-i\omega_0)\left(K_0^{(1)}+K_0^{(2)}+\Xi_{0,0}\right)}}{\gamma_i-i(\omega_{-1}-\omega_s)-\dfrac{\alpha^2\eta_0^{-1}W_{p0}i\omega_s}{(\gamma_i-i\omega_0)\left(K_0^{(1)}+K_0^{(2)}+\Xi_{0,0}\right)}}. \quad (C1)$$

Under the condition $\varepsilon_1=\varepsilon_2$, $\mu_1=\mu_2$, $\gamma_i=0$, we obtain

$$\frac{\left(\dfrac{2K_0^{(1)}}{2K_0^{(1)}-\Xi_{0,0}}\right)\left(\dfrac{\omega_s}{\omega_0}\right)\dfrac{\alpha^2\eta_0^{-1}W_{p0}}{\left(2K_0^{(1)}+\Xi_{0,0}\right)}}{\left(\dfrac{\omega_s}{\omega_0}\right)\dfrac{\alpha^2\eta_0^{-1}W_{p0}}{\left(2K_0^{(1)}+\Xi_{0,0}\right)}-i(\omega_{-1}-\omega_s)}=\frac{\dfrac{\left(\dfrac{\omega_s}{\omega_0}\right)\alpha^2\eta_0^{-1}W_{p0}\dfrac{2\omega_0\varepsilon_0\varepsilon_1}{k_{z,0}^{(1)}}}{\left(\dfrac{2\omega_0\varepsilon_0\varepsilon_1}{k_{z,0}^{(1)}}\right)^2+\left(\dfrac{\eta_0^{-1}W_{p0}}{\omega_0}\right)^2}}{\dfrac{\left(\dfrac{\omega_s}{\omega_0}\right)\alpha^2\eta_0^{-1}W_{p0}}{\dfrac{2\omega_0\varepsilon_0\varepsilon_1}{k_{z,0}^{(1)}}+i\dfrac{\eta_0^{-1}W_{p0}}{\omega_0}}-i(\omega_{-1}-\omega_s)}. \quad (C2)$$

Plugging $k_{z,0}^{(1)}=\omega_0\sqrt{\varepsilon_1\mu_1}\cos\theta_i/c$ into (C1), we have

$$\frac{\left(\dfrac{\omega_s}{\omega_0}\right)\dfrac{2\alpha^2W_{p0}\eta_1^{-1}/\cos\theta_i}{\left(2\eta_1^{-1}/\cos\theta_i\right)^2+\left(W_{p0}/\omega_0\right)^2}}{\left(\dfrac{\omega_s}{\omega_0}\right)\dfrac{\alpha^2W_{p0}\left(2\eta_1^{-1}/\cos\theta_i-iW_{p0}/\omega_0\right)}{\left(2\eta_1^{-1}/\cos\theta_i\right)^2+\left(W_{p0}/\omega_0\right)^2}-i(\omega_{-1}-\omega_s)}=\frac{\gamma_s}{\gamma_s-i(\omega_{-1}-\omega_s')}, \quad (C3)$$

which proves Eq. (41)

**Acknowledgement**
The authors acknowledge financial support from the National Science and Technology Council under Grant No. NSTC 113-2112-M-110-007-MY3.